# covEcho – Resource constrained lung ultrasound image analysis tool for faster triaging and active learning

Jinu Joseph[1], Mahesh Raveendranatha Panicker[1], Yale Tung Chen[2], Kesavadas Chandrasekharan[3], Vimal Chacko Mondy[3], Anoop Ayyappan[3], Jineesh Valakkada[3] and Kiran Vishnu Narayan[4]

[1]Center for Computational Imaging, Indian Institute of Technology Palakkad, 678623 India
[2]Hospital Universitario Puerta de Hierro, Madrid 28222 Spain
[3]Sree Chitra Tirunal Institute for Medical Sciences & Technology, Thiruvananthapuram, 695011 India
[4]Government Medical College Thiruvananthapuram, 695011 India

*Abstract*—Lung ultrasound (LUS) is possibly the only medical imaging modality which could be used for continuous and periodic monitoring of the lung. This is extremely useful in tracking the lung manifestations either during the onset of lung infection or to track the effect of vaccination on lung as in pandemics such as COVID-19. There have been many attempts in automating the classification of severity of lung into various classes or automatic segmentation of various LUS landmarks and manifestations. However, all these approaches are based on training static machine learning models which require a significantly clinically annotated large dataset and are computationally heavy and most of the time non-real time. In this work, a real-time light weight active learning-based approach is presented for faster triaging in COVID-19 subjects in resource constrained settings. The tool, based on the you look only once (YOLO) network, has the capability of providing the quality of images based on the identification of various LUS landmarks, artefacts and manifestations, prediction of severity of lung infection, possibility of active learning based on the feedback from clinicians or on the image quality and a summarization of the significant frames which are having high severity of infection and high image quality for further analysis. The results show that the proposed tool has a mean average precision (mAP) of 66% at an Intersection over Union (IoU) threshold of 0.5 for the prediction of LUS landmarks. The 14MB lightweight YOLOv5s network achieves 123 FPS while running in a Quadro P4000 GPU. The tool is available for usage and analysis upon request from the authors.

I.    INTRODUCTION

LUNG ultrasound (LUS) has been gaining significance as an inevitable clinical tool for the quick triaging and continuous monitoring of various pulmonary diseases [1, 2]. LUS has the advantages of the reduced risk of radiation, simplicity, portability, repeatability, non-invasiveness, the flexibility of use in emergency medicine and ease of equipment disinfection when compared to the standard diagnosing techniques such as computed tomography (CT) and magnetic resonance imaging (MRI) [2]. Air, being a highly attenuating medium to diagnostic ultrasound frequencies of 1 MHz and above, could have ideally restricted the usage of LUS. However, the artefacts created by the LUS have been employed cleverly by the clinical community to gauge the health of the lung [1].

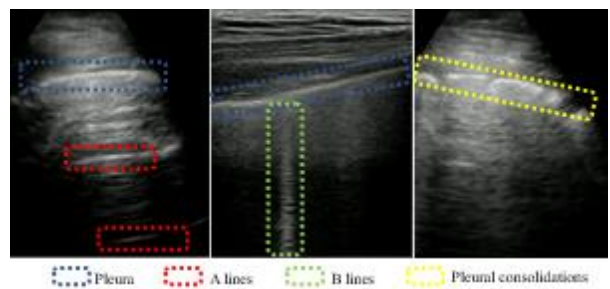

Fig. 1 Illustration of pleura, A lines, B lines and pleural consolidations in LUS imaging

A healthy lung is always characterized by a balanced ratio of tissue, air, and fluid. The large difference in the acoustic impedance between the visceral pleura (the membrane that covers the surface of each lung) and the air inside the alveolar space, gets manifested as a specular reflector that completely reflects the ultrasound (US) waves thus obstructing the penetration into further depths. However, the high-energy reflected US waves get reflected multiple times between the pleura and the transducer inducing reverberations [3]. The reverberations repeatedly appear as hyper-echoic horizontal lines of decreasing intensity beneath and parallel to the pleural line at equidistant intervals. The pattern is called the A-line which is a clinical representation of a healthy lung [3-5]. However, the presence of pulmonary diseases would alter the specular behavior of the pleura deterring the A lines and inducing newer acoustic artifacts such as B lines depending on the severity of infection. The B lines originate as discrete hyper-echoic vertical artifacts from a point along the pleural line and perpendicular to the line [6]. They are generated when there is an aeration reduction in the alveolar space due to the presence of extravascular lung water (EVLW) [7] or tissue consolidations [8]. Typical LUS images characterizing the important landmarks of pleura, and artefacts such as A lines and B lines are shown in Fig. 1 (Fig. 4 and 7 also has more details of the same).

Detection of pleura, A and B lines will be important for any approach to employ LUS for monitoring of lung infection. Several data-driven approaches for pleural segmentation and extraction of A, B lines have been proposed in the literature [9 - 12]. However, such architectures may be impractical for the current scenario of COVID-19, due to the minimal availability of training samples. Moreover, with COVID-19, spreading to remote locations, there may be a lack of trained clinicians who can perform US scans. Further, the clinical setting in hospitals may not be equipped with the latest US system to produce high quality scan images. Such situations, therefore, demand light weight real-time algorithms that detect the pleural line, A, B lines and pleural consolidations to aid the clinicians in the visual analysis and interpretation in real-time.

In this work, a lightweight image analysis tool with active learning capability is proposed. The approach is based on the you look only once (YOLO) framework [14] and can predict the quality of the image based on the detection of important LUS landmarks, predict the severity of the lung infection and also capable of retraining the model based on the predicted image quality or feedback from the clinicians. The rest of the paper is organized as follows. Section II discusses the proposed image analysis tool in detail. The detailed analysis and results on various applications of the proposed tool is presented in Section III. Section IV concludes the paper with possible directions for future work.

## II. PROPOSED IMAGE ANALYSIS TOOL

In this work, an image analysis tool named covEcho is presented, which could be employed in a resource constrained environment for real-time image analysis and feedback. The basic flow diagram of the tool is as shown in Fig. 2. Various capabilities of the tool are described in the subsequent sections.

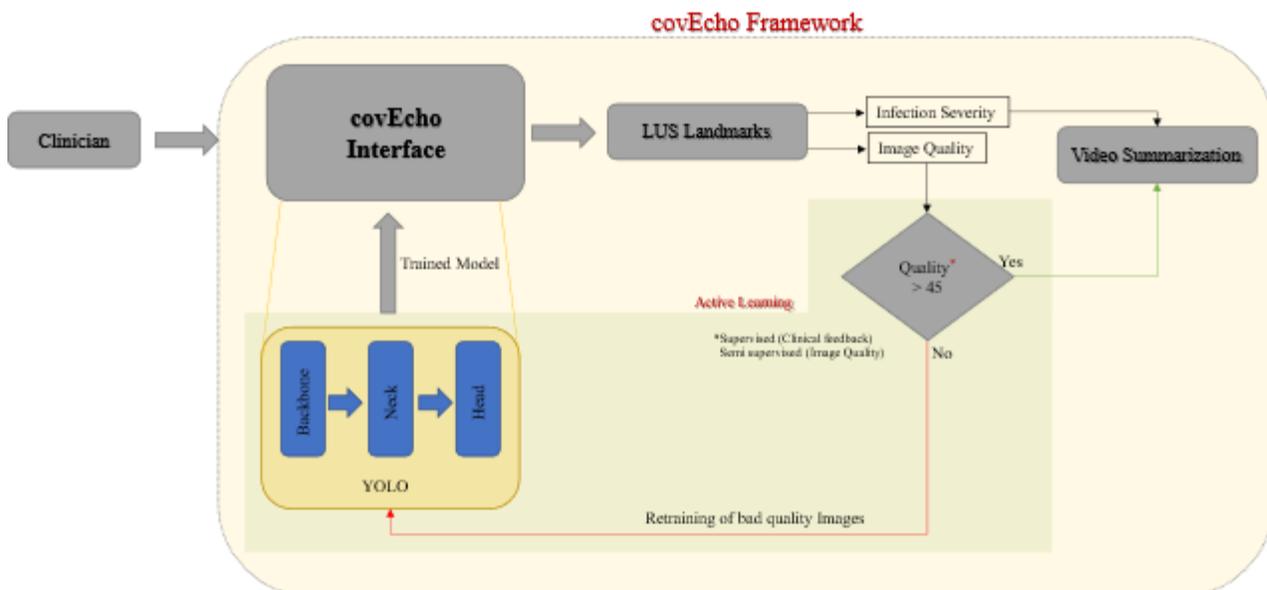

**Fig. 2. Proposed covEcho framework**

The basic building block of the proposed approach is the light weight YOLO neural network which will be trained on a limited dataset to identify the important landmarks in LUS images such as pleura, rib, and shadow, the artefacts such as A-lines or B-lines or B patches and the manifestations such as lung consolidations or air-bronchograms (see Fig. 4 and 7 for examples). Once this landmarks are identified, the output can be used for two purposes: 1) for checking a landmark driven image quality where the highest quality implies the essential landmarks such as pleura, rib and shadow and at least one of the artefacts or the manifestations are present and 2) for checking the infection severity of the lung, where a healthy lung implies the image with A-lines only and a most severe lung implies consolidations with air bronchograms. The interface provides the opportunity for active learning in a supervised or unsupervised way, where the supervised way is where the clinician overrides the labels which he feels is wrongly marked. In the unsupervised approach, the images with low image quality and also those with only pleural detections can be retrained subsequent to manual annotation. Also, based on the image quality and severity, a summary of the original LUS video with frames of high image quality and with presence of abnormal landmarks or manifestations can be made. This reduces the size of the videos (which aids telemedicine given the bandwidth constraints) as well as help the clinicians to take a relook at the significant frames in the video. The working of the proposed approach is as illustrated in the link and also available as a supplementary material.

The main contributions of this proposed image analysis tool can be enlisted as follows:

1) The covEcho tool offers a real time object detection on both LUS images and video frames and achieves appreciable results in terms of both speed and accuracy.
2) This lightweight YOLOv5s based tool can be easily implemented on embedded devices like Jetson Nano and can be used for clinical purposes in a resource constrained environment.
3) The object detection helps in unsupervised image quality and lung infection severity scores which could potentially have many applications.
4) The tool also supports a self- triggered relearning capability (including federated learning capabilities as in [25]), if the user is not satisfied with the detection results.
5) The tool also provides a summarized analysis report based on the network detection such that the clinicians get more assistance in diagnosis.

*A.    Data Acquisition*

The proposed framework makes use of a subset of 1200 lung ultrasound videos taken from 100 subjects using different ultrasound machines (GE Venue, Philips Lumify and Philips iU22) and over various geographies (India and Spain). The data were collected in association with clinicians from Hospital Universitario Puerta de Hierro, Majadahonda, Spain, Sree Chitra Tirunal Institute of Medical Sciences and Technology, Thiruvananthapuram, India, and Government Medical College, Kottayam, India. The details of the protocols followed at the various locations are as given below.

**Hospital Universitario Puerta de Hierro:** The ultrasound exam was performed with the patient in supine or near-supine position for the anterior scanning, and in the sitting or lateral decubitus position for the posterior scanning. The probe was positioned obliquely, along the intercostal spaces. The LUS examination was obtained moving the probe along anatomical reference lines (Fig. 12), 2nd-4th intercostal space (ICS) of parasternal, midclavicular, anterior axillary and midaxillary line (on the right side to the 5th ICS), whereas for the posterior chest, the paravertebral (2nd- 10th ICS), sub-scapular (7th- 10th ICS) and posterior axillary (2nd- 10th ICS) lines. A video clip was recorded along each anatomical line, recording at least 3 seconds at each ICS. The examinations were performed using a GE VENUE ultrasound system fitted with a phased and curvilinear array transducer (1.5–4.5 MHz) (courtesy General Electric Healthcare, Madrid, Spain).

**Sree Chitra Tirunal Institute of Medical Sciences and Technology:** The chest wall is divided into 12 imaging regions, 6 on each side - an anterior, lateral, and posterior zone, demarcated by the anterior and posterior axillary lines. Each region is then divided in two, superior and inferior. The patient is scanned in the supine position for the anterior and lateral zones and in the left lateral decubitus position for the posterior zones. The lung is scanned using a curvilinear transducer (C5-1 broadband curved array transducer) on Philips iU22 ultrasound machine in a longitudinal or sagittal orientation in each region to include the interposition between ribs and pleural line. Scanning is performed with the depth set at 8-10cm and the focus position at the pleural line. Video clips with a minimum of 100 frames (5 seconds of acquisition at 20 fps) are acquired in each region.

**Government Medical College**: Data acquisition was done using Lung preset setting of the Philips Lumify portable Ultrasound machine with 4-12 MHz frequency probe on patients with Severe and critical COVID 19 using a structured protocol as per lung ultrasound protocol suggested by Solidati et al. [3]. Longitudinal scans were done to rapidly screen the lungs and transverse scans were done to characterize severity of lung involvement.

*B.    Neural networks for object detection*

The major component of the covEcho tool is an object detection neural network. In computer vision, several object detection algorithms for detecting multiple object classes in images were developed which includes single stage detectors and two stage detectors. Single stage detectors treat object detection as a single regression problem without generating a sparse region of interest (RoI) set. On the other hand, two stage detectors generate sparse region proposals in the first stage and use the region proposals for object classification and bounding-box regression in the second stage. Two-stage detectors reach highest detection accuracy whereas single stage detectors take the lead in inference speed. In this work, we have considered only single stage detectors in LUS landmarking tasks in order to meet the real time requirements. Single Shot MultiBox Detector (SSD), You Only Look Once (YOLO), RetinaNet, EfficientDet are the leading networks in real time object detection tasks [14-18].

In 2015, Joseph Redmon et. al. introduced an object detection network named YOLO [14]. The main attraction behind this network was that YOLO transforms the object detection task as a single regression problem, such that the model predicts both bounding boxes and class probabilities directly from the image pixels. Subsequent to the initial version of YOLO, another two versions of YOLO were released that outperform the state of the art algorithms in terms of both speed and accuracy. Later YOLOv4 [19] and YOLOv5 [20] versions were also released which outperformed the previous versions in a significant way. The main difference between SSD and YOLO is that the SSD network utilizes convolutional layers of varying size, whereas YOLO architecture employs two fully connected layers for object detection.

YOLO computes all the features of the images and predicts all objects at the same time. As in paper [14], YOLOv1 applies a grid call of size NxN (7x7 default) into an image. The grid cell which contains the center of an object will be maintained and all other grid cells will be discarded. Each grid cell predicts B bounding boxes and its confidence scores. Each bounding box takes four values (x,y,w,h). where x,y corresponds to the center coordinates of the bounding box. w and h are the width and height of the bounding box respectively. The confidence score indicates the product of probability of the object inside the cell and intersection over union (IoU) of prediction box and ground truth box. For a dataset with a C number of classes, the total parameters are NxNx(5xB+C) [21]. YOLO uses a Non- Maximum Suppression (NMS) algorithm to remove all unwanted bounding boxes that do not contain any object and also to remove the bounding boxes that contain the same object as other bounding boxes. If the threshold value is less than the IoU of bounding boxes those boxes will be removed.

In this work, the latest YOLOv5s have been employed [20] for our experiment which has increased detection accuracy and speed compared to all other previous YOLO versions. It is natively implemented in PyTorch rather than Darknet. An overview of YOLOv5s model architecture is given in Fig. 3 [20, 22]. The network consists of three main components [20]:

- *Backbone:* Mainly used to extract input image features at different granularities. In YOLOv5 Cross Stage Partial Networks (CSP) Darknet as the backbone.
- *Neck:* A series of layers to process image features into semantical features and pass them forward to prediction.
- *Head:* Receives features from the neck and performs dense predictions.

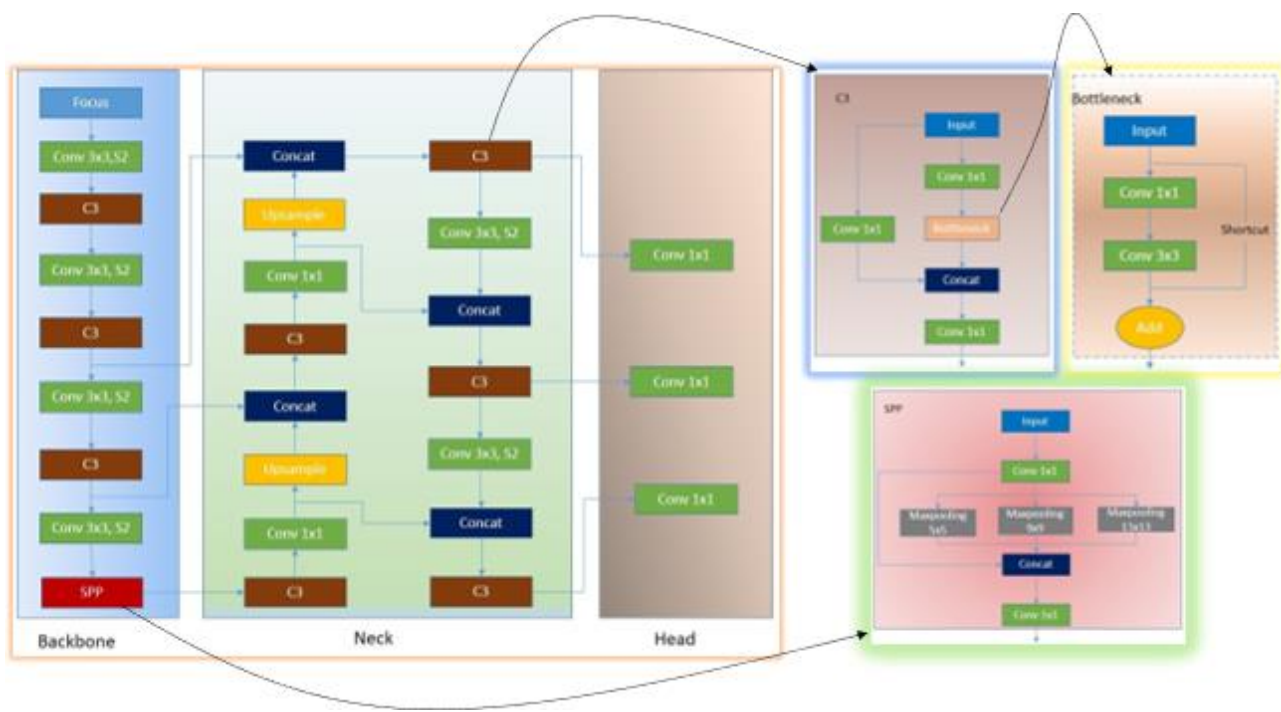

*Fig. 3 Yolov5 network architecture. It consists of 3 parts: (1) Backbone: CSPDarknet, (2) Neck: PANet, and (3) Detection Head. The input data given to CSPDarknet for feature extraction, and then fed to PANet for feature fusion. Finally, the detection head performs dense detection.*

Many object detectors make use of networks designed for image classification as backbones for extracting input image features.

The performance of object detectors largely depends on this backbone feature extractor. In YOLOv5 4.0 release the input feature extracted by backbone is sent to a spatial pyramid pooling (SPP) block to increase the receptive field and separate out the most important features by pooling multi-scale versions of it. In Fig. 3, Conv block (indicated with a kernel size of 3x3 or 1x1, S2 stands for stride of 2) consists of a Conv2D layer followed by a 2D batch normalization and sigmoid linear unit (SiLU) activation layer. Detection neck serves as a feature aggregator which mixes and combines the features received from the backbone and forward it to the detection head. The path aggregation network (PAN) architecture in the detection neck ensures a better small and medium-scale detector by concatenating the semantic features and fine-grained features at high-level layers. Detection head is responsible for performing localization and classification tasks. Unlike two stage detectors, one stage detectors implements these two tasks at the same time as a dense detection.

Single shot multiBox detector (SSD) [15] is a single shot detector that can directly predict the target category and position without any feature resampling and proposal generation as stated earlier. The architecture mainly constitutes a feature extraction part and a bounding box generation part. In the feature extraction part, SSD uses VGG16 [26] as the base network. The bounding box generation part makes use of 6 convolutional layers. The auxiliary convolution layers extract features at multiple scales. Each prediction contains a boundary box and C+1 scores, where C is the total number of classes and one for no object class. SSD loss calculation involves two components ie. confidence loss and localization loss. Confidence loss shows how confident the network is about the objectness of the predicted box. Localization loss shows the offset between predicted bounding box and ground truth bounding box.

RetinaNet [17] also has a one stage framework like YOLO and SSD composed of a backbone and two sub networks. RetinaNet architecture uses a feature pyramid network (FPN) backbone on top of a ResNet [27] architecture. The backbone of RetinaNet computes a convolutional feature map of the input image. Out of the two sub networks one subnet is for classifying anchor boxes and the second subnet is for generating offset from the anchor boxes to the ground truth bounding box. In order to handle the class imbalance problem in one stage detectors due to the dense sampling of anchor boxes, a focal loss function is introduced in it over cross entropy loss function. Focal loss helps to reduce the loss contribution from easily classified examples and learning can concentrate more on correcting misclassified cases.

EfficientDet [18] is also another one stage detector architecture which uses EffcientNet as its backbone with significantly high accuracy and efficiency across a wide spectrum of resource constraints. EfficientDet has a scalable detection architecture with a weighted bidirectional feature pyramid network (BiFPN) for fast and easy multi-scale feature fusion. Unlike conventional FPN, which has only a limited top - down information flow, this BiFPN offers better semantics by connecting low level to high level features and vice versa. Thus BiFPN works as a feature network with a bidirectional feature fusion mechanism. The fused features are given to class and box network to predict the object class and object bounding box. The detector also introduces a compound scaling method for object detection which jointly scales up all dimensions (like depth/width/resolution) of the backbone network, BiFPN feature network and class/box prediction network.

In this work, we have made a performance comparison on the above-mentioned networks in order to find out the most reliable one for LUS landmarking in a real time scenario.

### C.    LUS Landmark detection

Even though ultrasound has not been the preferred method for imaging the lung, the popularity has increased with the COVID19 primarily due to its bedside imaging capability, easier disinfection and more importantly the potential to use it for periodic monitoring of the lung [3, 24]. One of the major drawbacks for ultrasound is that it is highly operator dependent and interpretation is tough in many cases due to not so great image quality. This is where machine learning algorithms have played a significant role. There has been much work towards automation of the lung severity classification [9-12] or lung landmark segmentation [12], however there has been hardly any work dedicated to a tool for easier annotation and active learning of the above mentioned algorithms. In LUS, the important landmarks are pleura, rib and shadow, depending on whether the scan is done using convex or linear probes. LUS is also useful because of the presence of certain artifacts such as A-lines (in the case of a healthy lung) and B-lines or patches (in the case of an unhealthy lung). Lung manifestations such as consolidations and air bronchograms show the severity of the lung and are clearly evident in LUS images. Detection of these various landmarks, artefacts and manifestations using LUS has been attempted before [9, 13]. In [9], an SSD based framework was employed to detect the key landmarks and was done on swine models. In [13], claimed to be the first object detection approach in lung ultrasound on humans, a faster RCNN based approach is proposed to detect the key landmarks and artifacts such as normal pleura, irregular pleura, thick pleura, Alines, Coalescent B-lines, Separate B-lines and Consolidations in neonatal LUS. To the best of our knowledge, ours is the first YOLO based object detection approach for adult lung in the literature. Typical landmark detection results using a bounding box approach employing a YOLO algorithm is as shown in Fig. 4.

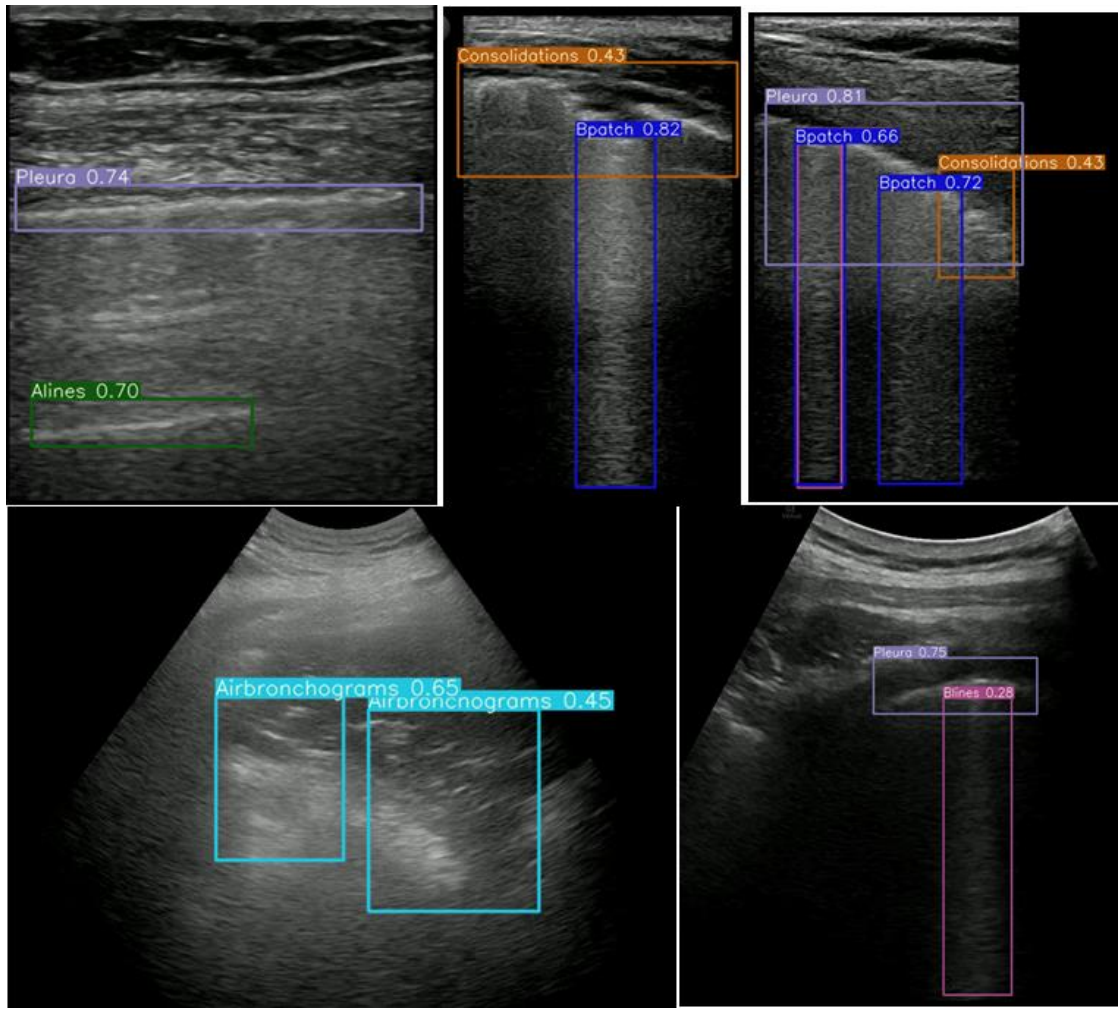

Fig. 4. Landmarks as detected by the YOLO approach

### D.  *Image Quality Measure*

One of the major issues with ultrasound images is the image quality in terms of the detection of the important landmarks, artifacts and manifestations in the LUS image. Towards this, a novel quality measure is proposed. Based on the sum of the quality scores of all landmarks in an LUS frame, covEcho tags each image with one of the following quality measures. ie. Excellent, Good, Average, Below average and Bad. As stated earlier, pleura is the most important landmark in a LUS image, which is independent of whether the scan is done using convex or linear probes. A quality score of 30 will be awarded to an image when the tool is able to detect at least pleura from it. For each separate identification of rib and shadow a quality score of 15 and 10 respectively will be given. If the tool is able to identify any one of the following artifacts ie. either A lines, B lines, B patch, Consolidations or Air bronchograms then the awarded quality score will be 45. Thus a total score of 100 will be given to an image with a pleura, rib, shadow and at least one of the above stated artifacts identified LUS scans. Fig. 5 shows the flowchart of quality measure marking in CovEcho annotated image.

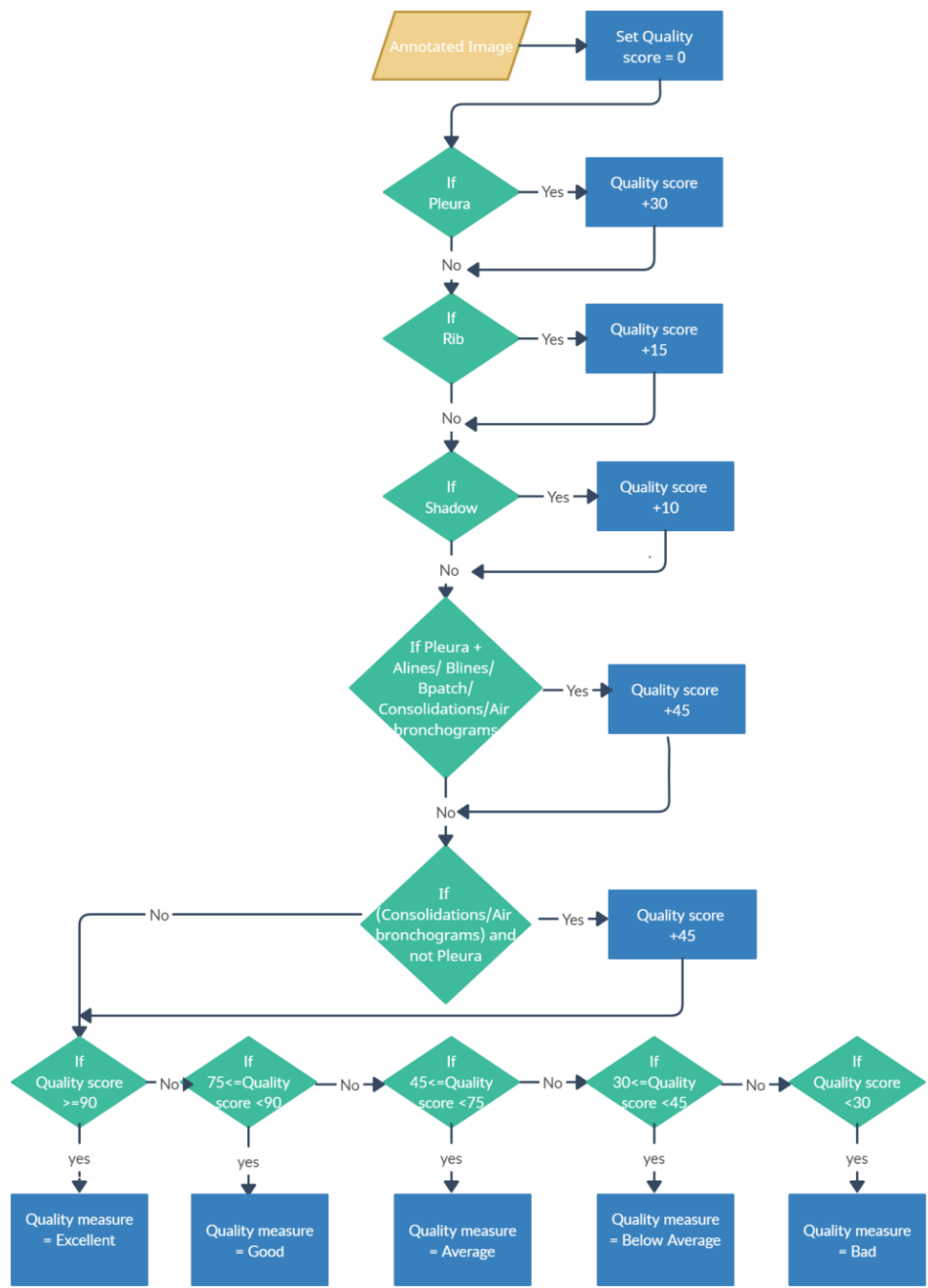

Fig.5 Flowchart of quality measure marking in covEcho annotated image.

An image with a total quality score greater than or equal to 90 will be considered as Excellent and if the total quality score greater than or equal to 75 but less than 90 then that image will be considered as Good. Similarly, a total quality score falls in the interval {45,75} will be taken as Average quality image. An image with a quality score in the interval {30,45} will be considered as below average one. Finally, an image with a total quality score less than 30 will be considered as a bad or low quality image. In Fig. 6, some annotated images with quality measures as Excellent, Good, Average, Below average and Bad respectively are shown.

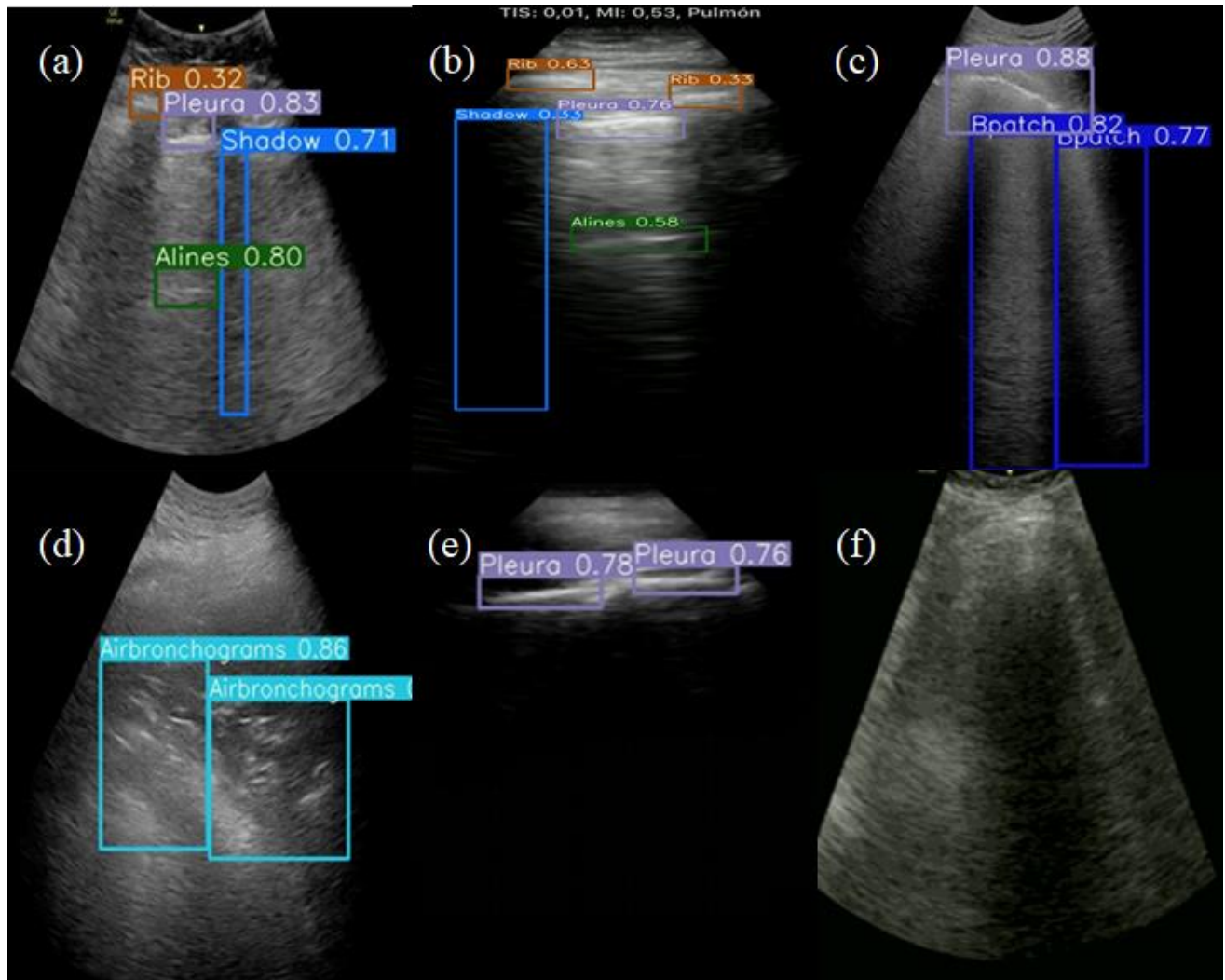

Fig. 6 shows different quality images based on covEcho annotations. a,b) An excellent quality image showing a healthy lung, c) A good quality image taken with a curvilinear probe, d) An average quality image with air bronchograms, e) A below average quality image pleura only and f) A bad quality undetected image.

### E.    Infection Severity

The severity classes of COVID-19 are decided from lung manifestations during disease progression as discussed in [4]. It includes the presence of A-lines (Class1 in Fig. 7), appearance of single or multiple B-lines (Class2 in Fig. 7) to a confluent appearance of B-lines (Class3 in Fig. 7), further degraded by the appearance of consolidations (Class 4 in Fig. 7) due to the effusion in between two pleural surfaces with or without the appearance of air bronchograms (Class 5 in Fig. 7).

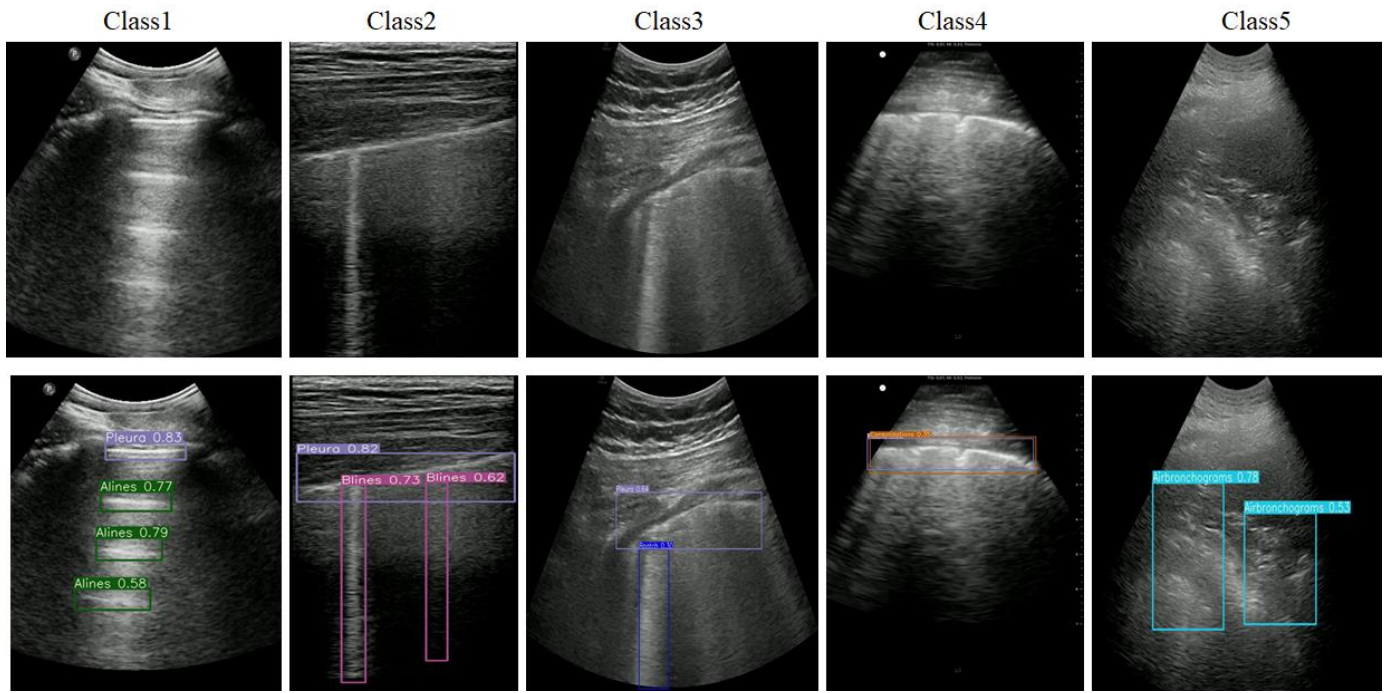

Fig. 7 Illustration of various classes of infection severity of the lung (top row) and the landmarks as detected by the covEcho tool (bottom row)

covEcho uses a severity score measure to mark each LUS frame to a particular class. We have considered pleura as a mandatory landmark in severity score decision and class identification. A severity score of 0, 1, 2, 3 and 4 are given to frames with A-lines, B-lines, B-patch, Consolidation and Air bronchograms respectively. The highest severity score will be taken as the severity score of a frame or an image. For a given LUS video, the tool calculates a severity score for each of the frames in the video. The overall severity score of the video is also decided as the highest severity score out of all frames in that particular video ie. The severity score for the video will be the severity score of the most severe frame in that video. The algorithm for the same is illustrated as in Fig. 8. For all undetected frames the severity score will be taken as -2 and mark those frames to Class 0. Within the detected frames, if any of the above mentioned lung manifestations are not identified the severity score of those frames will be taken as -1 and put those frames to Class6. As depicted in Fig. 7, frames with a severity score of 0, 1, 2, 3 and 4 will be categorized to Class 1, Class 2, Class 3, Class 4 and Class 5 respectively.

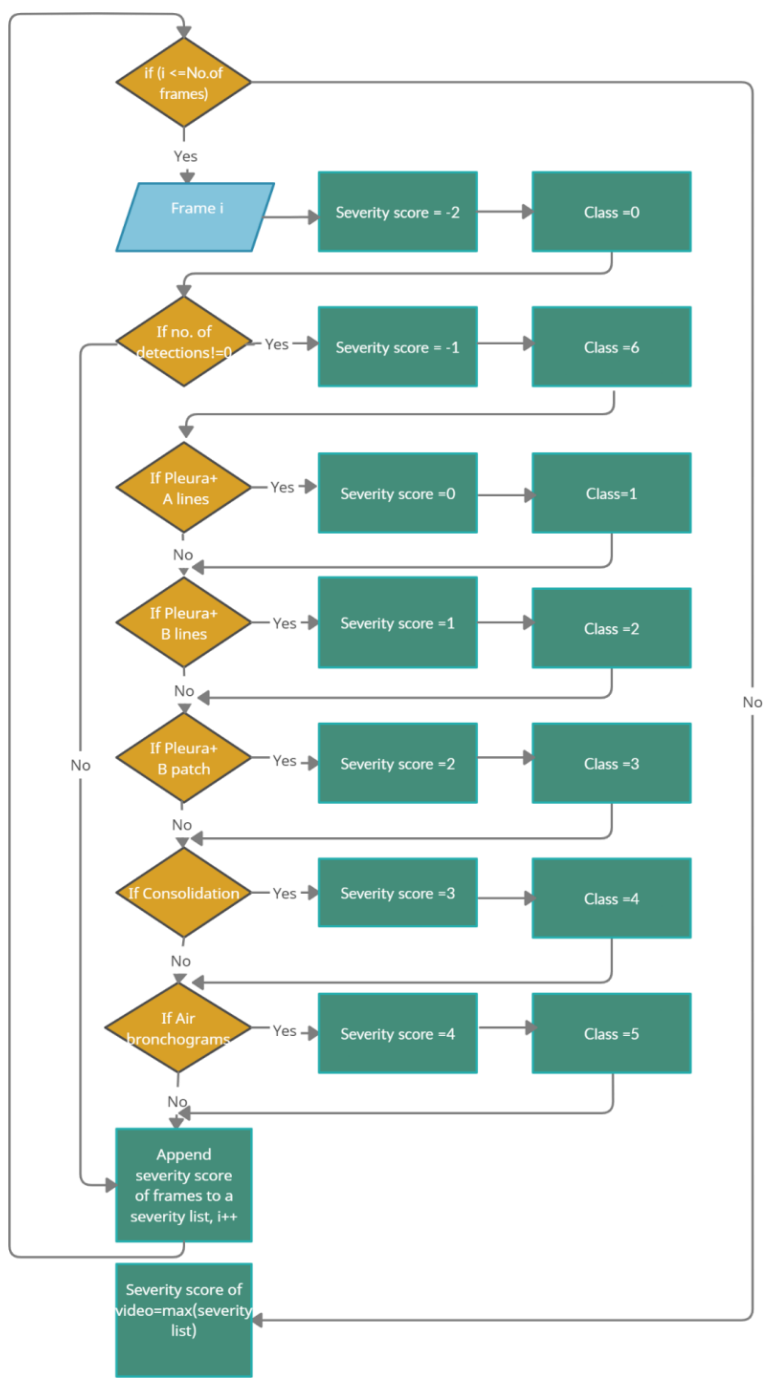

Fig. 8 Flowchart of severity score decision making and class identification in covEcho

## III. RESULTS

In this section, the results of the proposed tool in tasks of object detection, unsupervised image quality prediction, unsupervised severity classification both in terms of individual frames as well as for the video and the unsupervised video summarization tasks are presented.

### A.    *Landmark detection performance*

In the LUS landmark detection process, a balance of both accuracy and speed is required inorder to meet the real time requirements and hence the selection of an appropriate neural network is crucial. An ablation study has been conducted to compare the detection performance of single stage detectors like SSD, YOLOv5s, RetinaNet and EfficientDet in LUS object annotation task. Since all of them were conventionally trained on natural images, each of the above networks were completely retrained using LUS frames and compared in terms of both speed and accuracy in an identical testing environment.

The training was done employing a GPU cluster with eight A100 GPUs. Table 1 shows the hyperparameters used for training various object detectors. The training and testing was done employing a new custom dataset comprising 570 images for training and 163 images for validation. The performance comparison is made on 83 test images. All LUS images are labelled using labelImg [23], a graphical image annotation tool. The object detectors are trained for finding the 8 landmarks in a LUS frame irrespective of the probe shape as mentioned above. The distribution of each of the landmarks in the training dataset is given in Fig. 9. As expected, the dataset is dominated by pleura followed by B lines, which shows that the distribution is dominated by frames from abnormal lungs.

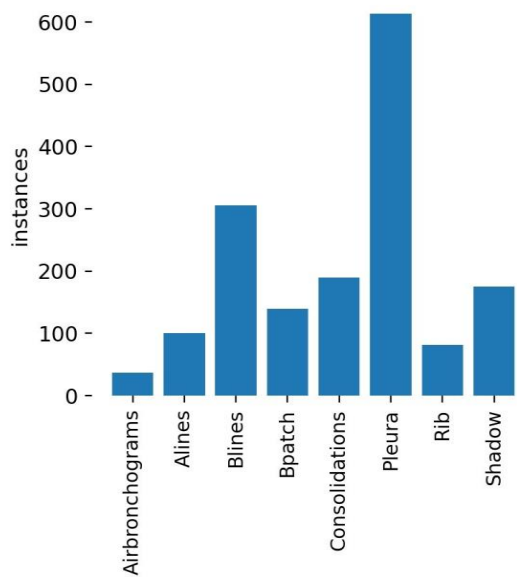

Fig. 9 The distribution of LUS landmarks in the training dataset.

Table 2 shows the summary of performance results of different object detectors. The second column indicates the backbone of the model. The third column represents the test dataset used. The fourth column is the mean average precision (mAP) in measuring accuracy for IoU=0.5. The fifth column indicates mean average precision where IoU varies from 0.5 to 0.95 with a step size of 0.05. The last column shows the timing on a Quadro P 4000 GPU. The mAP is measured with the 83 test images. For SSD, Table 2 shows results for 300x300 input images. For YOLO and for all other networks, the results are for 416x416 input images.

Table 1: The hyperparameters used for training one stage object detectors. lr0 and lf indicate initial and final learning rate respectively.

| Object Detector | Input resolution | Batch size | Learning rate(lr) | Number of epochs |
|---|---|---|---|---|
| YOLOv5s | 416x416 | 16 | lr0=0.01, lrf=0.2 | 300 |
| SSD | 300x300 | 8 | 0.0001 | 120 |
| RetinaNet | 608x608 | 2 | 0.00001 | 100 |
| EffcientDet | 512x512 | 8 | 0.0001 | 500 |

Table 2: The mean Average Precision Scores on the LUS test data for multiple detectors.

*This is an originally submitted version and has not been reviewed by independent peers*

| Object Detector | Backbone | Test data | mAP @0.5 | mAP @0.5:0.95 | FPS | Model Size (MB) |
|---|---|---|---|---|---|---|
| **YOLOv5s** | **CSPDarknet** | **83** | **0.660** | **0.293** | **123** | **14** |
| SSD | VGG -16 | 83 | 0.630 | 0.277 | 31 | 96 |
| RetinaNet | Resnet-50 | 83 | 0.529 | 0.247 | 4 | 92 |
| EfficientDet | EfficientNet | 83 | 0.357 | 0.154 | 17 | 17 |

The evaluation results shows that YOLOv5s on CSPDarknet backbone has the highest mAP among the other detectors targeted for real time processing. YOLOv5s has a good balance between accuracy and speed. SSD also attains a similar performance in terms of mean average precision. We haven't performed evaluation on SSD-512 as it will be slower than SSD-300. It should be noted that SSD requires an inference time of at least 32 ms per image whereas YOLOv5s requires only 8.1 ms. Even Though SSD and RetinaNet have reasonable mAP, YOLOv5s outperforms these models in terms of both speed and accuracy. Table 3 shows the class-wise performance of all detectors considered. In Table 3, YOLOv5s and SSD offer decent results for all LUS landmarks. RetinaNet fails to pick up B lines in detection. SSD has problems in detecting ribs compared to all other classes. Table 3 shows that YOLO is the most accurate model in LUS landmarking applications. Given that it is very light weight (14 MB), YOLO based approach is the ideal approach for resource constrained environments. Table 4 depicts EfficientDet average precision scores on the LUS test data for small, medium and large objects.

Table 3: LUS test data class wise results for multiple detectors.

| Class | YOLO mAP@0.5 | YOLO mAP@0.5:0.95 | SSD mAP@0.5 | SSD mAP@0.5:0.95 | RetinaNet mAP@0.5 | RetinaNet mAp@0.5:0.95 |
|---|---|---|---|---|---|---|
| Air bronchograms | 0.995 | 0.597 | 1 | 0.473 | 1 | 0.65 |
| A lines | 0.501 | 0.108 | 0.452 | 0.187 | 0.558 | 0.182 |
| B lines | 0.651 | 0.302 | 0.495 | 0.185 | 0.072 | 0.024 |
| B patch | 0.716 | 0.404 | 0.694 | 0.394 | 0.553 | 0.329 |
| Consolidations | 0.603 | 0.245 | 0.560 | 0.220 | 0.664 | 0.291 |
| Pleura | 0.771 | 0.322 | 0.762 | 0.350 | 0.713 | 0.322 |
| Rib | 0.589 | 0.177 | 0.344 | 0.099 | 0.238 | 0.052 |
| Shadow | 0.456 | 0.187 | 0.733 | 0.306 | 0.434 | 0.124 |

Table 4: EfficientDet Average Precision scores on the LUS test data for small, medium and large objects.

| IoU@ | Area | Max Dets | EfficientDet Average Precision |
|---|---|---|---|
| 0.5:0.95 | all | 100 | 0.154 |
| 0.5 | all | 100 | 0.357 |
| 0.75 | all | 100 | 0.107 |

| 0.5:0.95 | small | 100 | 0.143 |
| 0.5:0.95 | medium | 100 | 0.103 |
| 0.5:0.95 | large | 100 | 0.262 |

The class wise precision and recall of LUS landmarks for YOLOv5s is depicted in Table 5. The network has 224 layers with almost 7.5 million parameters. The model takes 5.6ms for inference, 2.5 ms for non max suppression resulting in a total of 8.1 ms for a 416x416 image at batch size 32. We have used the weight at the last epoch for the highest accuracy. The results show that the proposed tool has a sensitivity of 68.5% and precision of 67% for the prediction of LUS landmarks training at a confidence score threshold of 0.25.

Table 5: Class wise precision and recall of all LUS landmarks for YOLOv5 model.

| Class | Test Images | Labels | Precision | Recall |
|---|---|---|---|---|
| all | 83 | 233 | 0.67 | 0.685 |
| Air bronchograms | 83 | 2 | 0.807 | 1 |
| A lines | 83 | 12 | 0.637 | 0.5 |
| B lines | 83 | 43 | 0.668 | 0.581 |
| B patch | 83 | 21 | 0.653 | 0.762 |
| Consolidations | 83 | 35 | 0.642 | 0.665 |
| Pleura | 83 | 90 | 0.764 | 0.883 |
| Rib | 83 | 7 | 0.667 | 0.571 |
| Shadow | 83 | 23 | 0.519 | 0.565 |

Fig. 10 shows the precision, recall, F1 curves for test data inYOLO5s model. As we are more interested in lightweight networks for working in resource constrained devices, YOLOv5s has the smallest footprint. It takes less than 15 MB memory while SSD and RetinaNet requires more than 90 MB space. Hence, the lightweight YOLO based network can be implemented in embedded devices like the NVIDIA Jetson Nano module for clinical purposes.

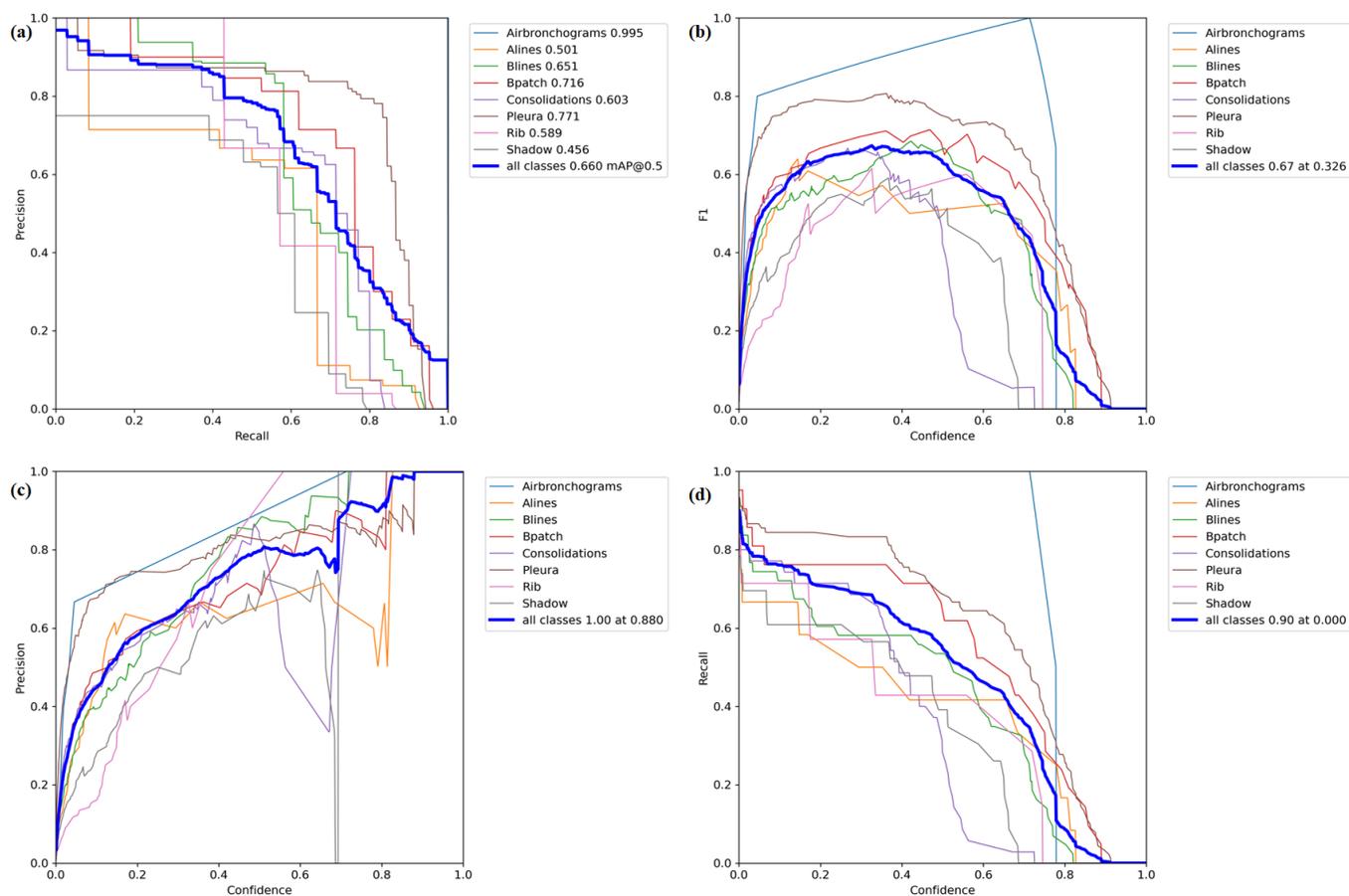

Fig. 10 a) Precision - recall curve for 8 object classes, where IoU threshold is 0.5. b) F1 vs Confidence score c) Precision vs Confidence score d) Recall vs Confidence score for 8 object classes for YOLOv5s model.

### B. *Unsupervised Image Quality*

As discussed, the proposed tool provides an image quality measure. To check the ability of the proposed approach to identify low quality frames, an ablation study has been conducted on a total of 600 images randomly selected from 1200 LUS videos. In these, around 200 frames were annotated as of bad quality by the clinicians. The covEcho tool was able to detect the image quality with a sensitivity of 95% and specificity of 97%. Fig. 11 shows some of the images which were selected as of bad quality (top row) and good quality (bottom row) as per the algorithm proposed in section 2.D. It is very evident that, in the case of bad quality images, even the pleural line is not very evident.

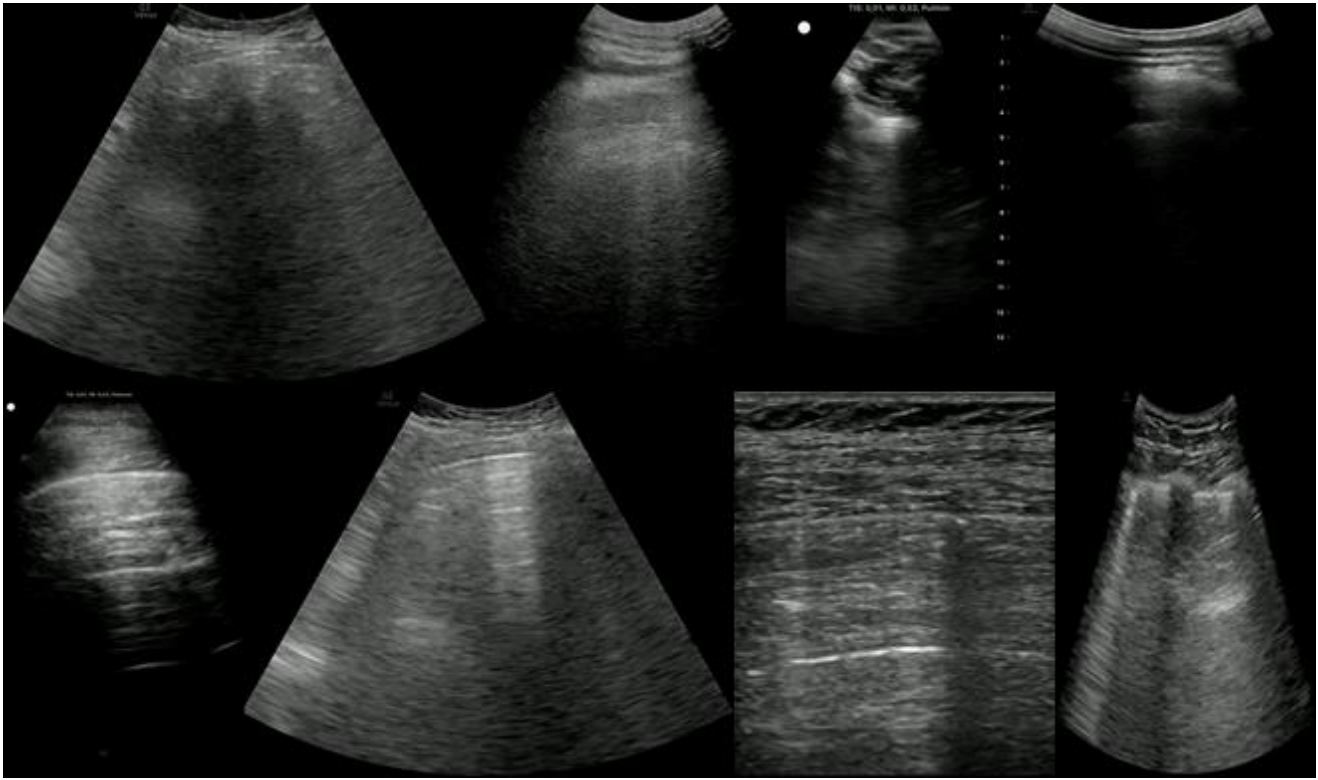

Fig. 11 Images of bad quality as selected by covEcho tool (top row) and Images of good quality as selected by covEcho tool (bottom row)

### C.    Unsupervised Classification

Since the covEcho tool estimates the infection severity and assigns a score to each of the frames in the video, it is possible to employ the same for classifying the frames and subsequently the video itself which will be extremely beneficial for diagnosis. The details of the estimation of infection severity are already explained in section II.E. In a specific study on the unsupervised classification, about 130 videos are randomly selected from the 1200 videos. Out of these 130 videos, 5495 frames are selected and annotated by clinicians into the five classes with image strength in each of the five classes are 1749, 896, 1428, 1253 and 169 respectively. It has to be noted that the frames belonging to class 5 with air bronchograms are very rare. The dataset has been employed to check the efficacy of the proposed object detection tool for a frame level classification performance evaluation. The results of the unsupervised classification are as shown in Tables 6 and 7. Table 6 shows the confusion matrix and Table 7 has the details of the performance of the classification. The proposed approach even without any supervised training achieves very good classification accuracy (ACC), sensitivity (SEN) and specificity (SPEC). It has to be noted that, the no class (which is due to only pleura detection in those images) are not considered for the calculation of accuracy, sensitivity and specificity.

Table 6: Class confusion matrix

| Actual Predicted | Class 1 | Class2 | Class3 | Class4 | Class5 | No class |
|---|---|---|---|---|---|---|
| Class1 | 1747 | 0 | 2 | 0 | 0 | 0 |
| Class2 | 37 | 831 | 28 | 0 | 0 | 29 |

*This is an originally submitted version and has not been reviewed by independent peers*

Jinu *et al.*: covEcho – Resource constrained lung ultrasound image analysis tool for faster triaging and active learning in COVID-19    7| | | | | | | |
|---|---|---|---|---|---|---|
| Class3 | 28 | 0 | 1395 | 0 | 0 | 5 |
| Class4 | 1 | 0 | 0 | 1233 | 0 | 19 |
| Class5 | 0 | 0 | 0 | 0 | 169 | 0 |

Table 7: Performance Analysis of the unsupervised classification approach

| Class | ACC | SEN | SPEC |
|---|---|---|---|
| 1 | 0.9876 | 0.9989 | 0.9823 |
| 2 | 0.9881 | 0.9275 | 1.0 |
| 3 | 0.9894 | 0.9803 | 0.9926 |
| 4 | 0.9998 | 0.9992 | 1.0 |
| 5 | 1.0 | 1.0 | 1.0 |

A video wise performance analysis of the covEcho tool is also done with the above mentioned 130 videos. Each of these videos are labelled as 0 for a normal lung and 1 for an abnormal lung. With reference to Fig. 8, we have considered the severity score of the worst detected frame from the video for marking it as normal or abnormal lung. If the severity score of the worst detected frame is greater than or equal to one, those videos will be put in the abnormal lung category. If the severity score of the video is equal to zero, those videos will be marked to normal category. If the severity score is less than zero, which means the tool either fails to pick any lung manifestations from the video or the detected frames contain only pleura/rib/shadow. Table 8 shows the unsupervised binary video classification performance of the CovEcho tool. The table shows the tool has an accuracy of 90.77% with a precision of 93.68% and a recall of 95.67%.

Table 8: Unsupervised binary video classification performance of the CovEcho tool.

| Groundtruth/ Prediction | Abnormal lung (1) | Normal lung (0) | Undetected |
|---|---|---|---|
| Abnormal lung (1) | 89 | 3 | 1 |

*This is an originally submitted version and has not been reviewed by independent peers*

| | | | |
|---|---|---|---|
| Normal lung (0) | 6 | 29 | 2 |

### D.     Proposed Video Summarization

It is very evident that LUS has been employed as a periodic monitoring tool for lung monitoring during COVID -19 [24]. During periodic monitoring, analysing each frame in a video corresponding to a particular location will be really a tedious procedure and most of the time impractical. So rather than going for each video frame by frame, it is a good practice to choose a single view which can encapsulate the outline of analysis. Towards this a data acquisition protocol for covEcho image analysis tool is employed as part of this work. LUS videos were acquired using continuous sweep of each rib space: seven at the right, seven at the left resulting in a total of 14 points from 36 subjects resulting in a total of 504 videos. A pictorial representation of the acquisition points and the sweep area is as shown in Fig. 12.

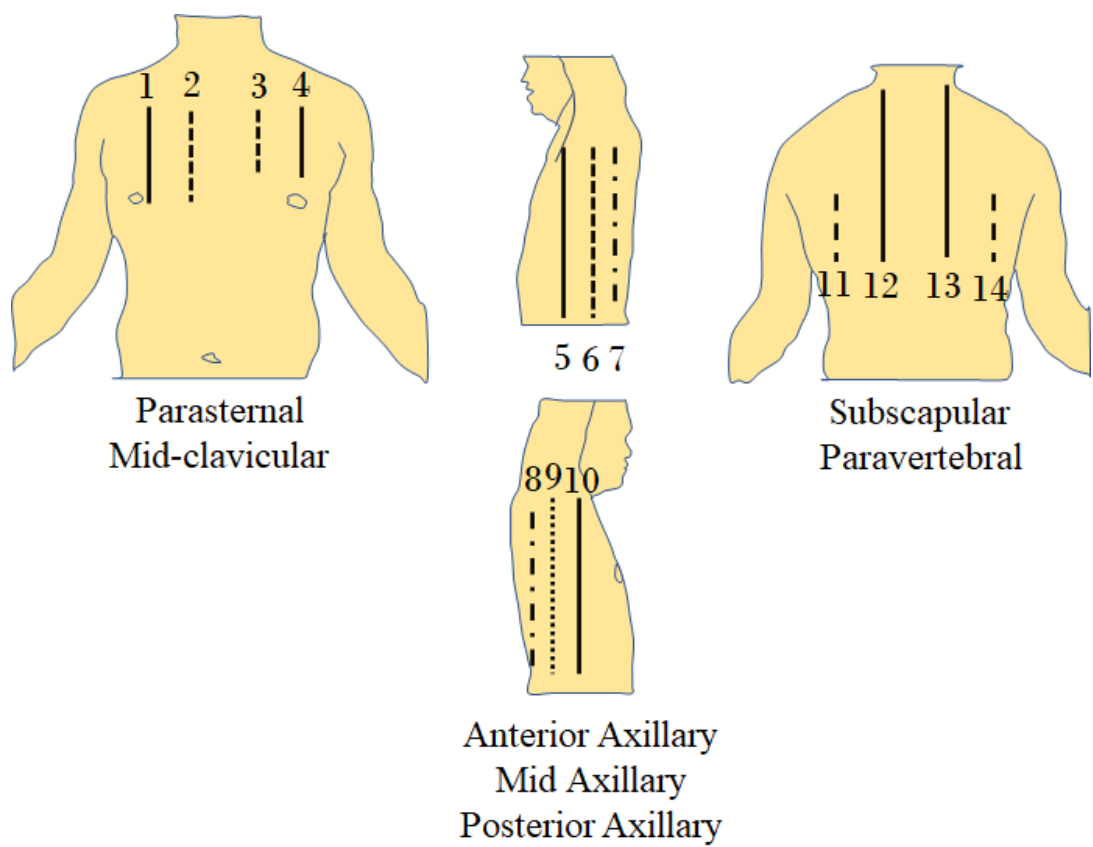

Fig. 12   Data acquisition protocol

Fig. 13 corresponds to a covEcho generated 14 point scan analysis view which encodes the severity classes using a colormap. For all 14 videos in the dataset, the tool finds a severity score which is then encoded as colormaps such that the clinicians can easily understand the progress of a patient day by day. In Fig. 13, the green color box corresponds to a zero severity score. Yellow - green color box corresponds to a severity score of 1. Similarly, yellow, orange color corresponds to a severity score of 2 and 3 respectively. Finally, the red color indicates the most severe class with a severity score of 4. If the data at a particular location is unavailable or any lung manifestations are not identified other than pleura, rib and shadow then it will be marked by a black color box (which is also visible in Fig. 13). The colormap encoding procedure helps to identify the worst case frame in each video corresponding to a location. The proposed covEcho tool also provides a box plot indicating the severity classes of all detected frames from the all 14 lung locations, which helps the clinicians in identifying the quantum of severity at a particular location. In a nutshell, the box plot helps the clinicians to get an overall idea about the severity classes of all frames. In some of the cases, the frames with the worst severity score (which essentially forms the part of the summary frames) are also shown in Fig. 13.

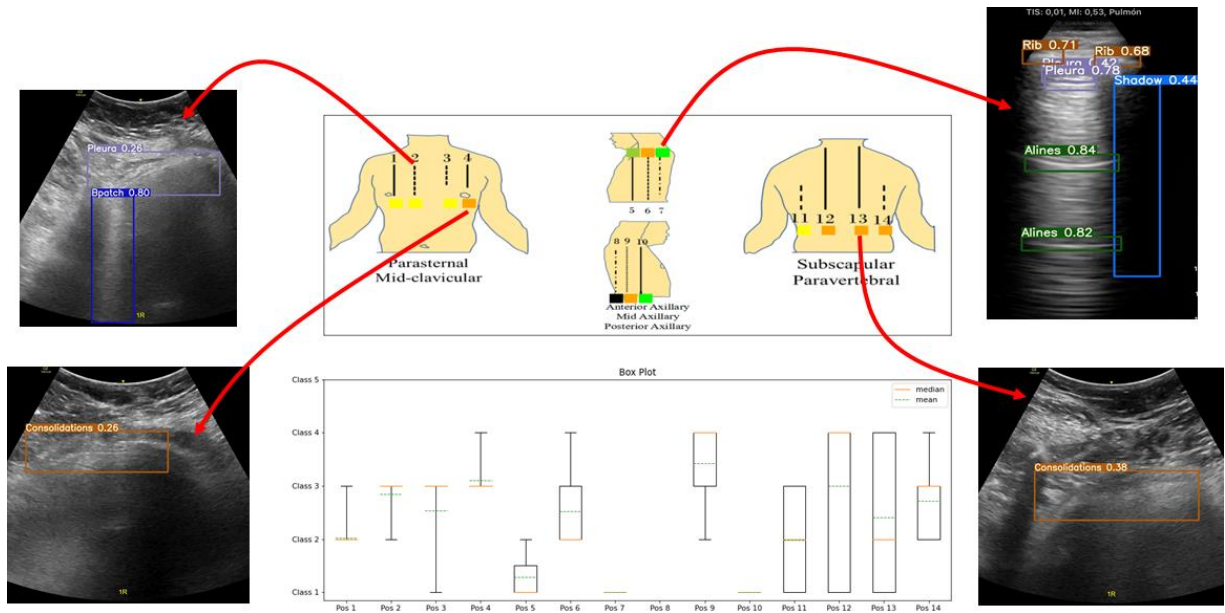

Fig. 13 covEcho generated a 14 point scan analysis colour code map

In addition to this covEcho does a video summarization in an unsupervised way, making the clinical evaluation fast. In the unsupervised video summarization, the tool collects all frames in the video which are having a severity score of 1 and above. The Fig. 14 shows covEcho identified the most severe frames in different videos.

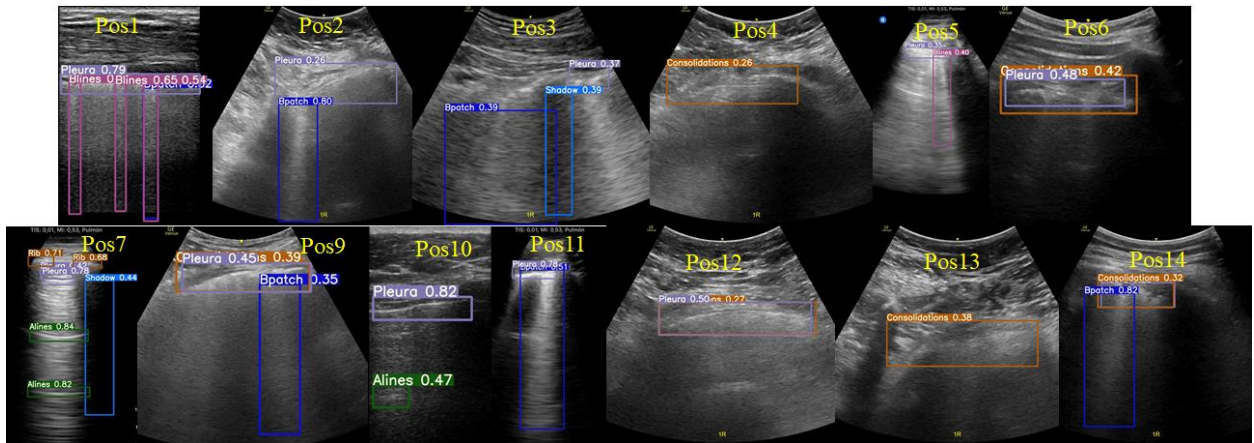

Fig. 14 covEcho selected the most severe frames from unsupervised video summarization for the case in Fig. 13

## IV.    CONCLUSIONS

With the onset of COVID-19, lung ultrasound has been employed as a powerful tool for continuous monitoring of the lung. There have been many works reported in literature towards automatic severity classification and landmark detection in the lung. However all such approaches have been based on static models which require a significant amount of diverse annotated dataset. In this work, a real-time light weight active learning-based approach is presented for faster triaging in COVID-19 subjects in resource constrained settings. The tool, based on the You Look Only Once (YOLO) network, has the capability of providing the quality of images based on the identification of various LUS landmarks, artefacts and manifestations. Prediction of severity of lung infection, possibility of active learning based on the feedback from clinicians or on the image quality and a summarization of the significant frames which are having high severity of infection and high image quality for further analysis are the added advantages of this tool. The performance analysis shows that the proposed tool has a mean average precision (mAP) of 66% at an Intersection over Union (IoU) threshold of 0.5 for the prediction of LUS landmarks. One of the issues with the proposed work has been the wide variance in the type of data acquired. Future work involves introducing some pre-processing to standardize the ultrasound images or frames before employing the same for analysis. Also, it is envisaged to develop a hardware package which includes a low-cost edge processor (e.g. NVIDIA Jetson Nano) to be connected to ultrasound systems where the covEcho tool can be deployed.

SUPPLEMENTARY MATERIALS

The detailed video demo of the covEcho tool is attached as supplementary files and are available online. The executables will be made available upon request under license.

ACKNOWLEDGMENT

The authors would like to acknowledge the funding from the Department of Science and Technology - Science and Engineering Research Board (DSTSERB (CVD/2020/000221)) for the CRG COVID19 funding. The authors are grateful to Mr. Jigar Halani, Dr. Manish Modani, Mr. Megh Makwana and Mr. Prakash Tubakad (all of them from NVIDIA) for providing adequate help in getting access to the CDAC PARAM SIDDHI AI system. We are extremely thankful to the CDAC team for giving access to the PARAM SIDDHI AI system without which the project wouldn't have been completed on time.